\newcommand{\Deqn}[1]{{Eq.~(\ref{#1})}}

\newcommand{\Dfig}[1]{{Fig.~\ref{#1}}} \newcommand{\beq}{\begin{equation}}
\newcommand{\eeq}{\end{equation}} \newcommand{\bea}{\begin{eqnarray}}
\newcommand{\eea}{\end{eqnarray}}

\documentclass[twocolumn,prd,aps,amssymb,nofootinbib,epsf]{revtex4}
\usepackage[usenames]{color}
\usepackage{bm} \usepackage{amsmath} \usepackage{amsfonts} 
\usepackage{graphicx}

\begin{document}

\title{Gravitational Waves from Global Second Order Phase Transitions}

\author{John T. Giblin, Jr${}^{1,2,3}$} \author{Larry R. Price${}^{4,5}$}
\author{Xavier Siemens${}^4$} \author{Brian Vlcek${}^4$}

\affiliation{${}^1$Department of Physics, Kenyon College, Gambier, OH 43022}
\affiliation{${}^2$Perimeter Institute for Theoretical Physics, 31 Caroline St.
N, Waterloo, ON N2L 2Y5} 
\affiliation{${}^3$Department of Physics, Case Western
Reserve University, Cleveland, OH 44106} 
\affiliation{${}^4$Center for
Gravitation and Cosmology, Department of Physics, University of
Wisconsin--Milwaukee, P.O. Box 413, Milwaukee, WI  53201} 
\affiliation{${}^5$ LIGO -- California Institute of Technology, Pasadena, CA
91125, USA}
\date{\today}

\begin{abstract} Global second-order phase transitions are expected to produce
scale-invariant gravitational wave spectra.  In this manuscript we explore the
dynamics of a symmetry-breaking phase transition using lattice simulations.  We
explicitly calculate the stochastic gravitational wave background produced
during the transition and subsequent self-ordering phase.  We comment on this signal
as it compares to the scale-invariant spectrum produced during inflation.
\end{abstract}


\maketitle

\section{Introduction}

A scale-invariant spectrum of gravitational radiation is a key prediction of
inflation \cite{Starobinsky:1979ty,Rubakov:1982df}.  Measuring the ratio of the
amplitude of gravitational radiation to the amplitude of density perturbations,
$r$, would be a direct probe of the inflationary energy scale.  Such a
measurement has eluded observation in the cosmic microwave background (CMB) thus
far \cite{Komatsu:2010fb} and represents one of the key goals of future CMB
observational missions \cite{:2006uk}.

On the other hand, some authors have argued that phase transitions
can mimic the scale-invariant inflationary signal
\cite{Krauss:1991qu,JonesSmith:2007ne,Fenu:2009qf}.  The mechanism through which
this is accomplished is not trivial.  The phase transition itself is not the
source. Rather, energy is deposited into gravitational radiation via the
self-ordering of fields as regions of spacetime become causally connected.  

The process begins with $\mathcal{N}$ scalar fields, $\phi_i$, subject to a temperature
dependent potential, $V$.  At high temperatures the potential looks
quadratic with a single minimum
at the origin.  This state has $O(\mathcal{N})$ symmetry since the state is
symmetric any any rotation in the $\mathcal{N}$-dimensional field space.  Once
the potential drops below a critical temperature, the field acquires a vacuum
expectation value (VEV).  This process spontaneously breaks the $O(\mathcal{N)}$
to a $O(\mathcal{N}-1)$ symmetry.  The transition can be very fast, with a
duration, $\tau$, that is much smaller than the Hubble time, $\tau \ll
H^{-1}$.  The field configuration of each Hubble volume will lie somewhere in
the vacuum state, however each causally disconnected volume will be independent of the others.  This
process can be quite smooth, and there is no {\sl a priori} reason to believe
such a process radiates.
However, if the Universe is radiation (and later matter) dominated after the transition, the growth
of the comoving Hubble scale continuously brings previously causally disconnected regions into
contact.  Regions thus acquire field gradients on the scale of the Hubble
horizon.  These gradients generate anisotropic stresses that source
gravitational waves.  The growth of the horizon acts as a high-pass filter on the
spectrum of gravitational waves, {\sl freezing out} large-wavelength 
modes until they enter the horizon and become dynamical.
As the horizon grows power is distributed at all scales between the physical
Horizon size at the time of the transition, $H_c^{-1}$, and the Horizon size at
the time of observation.  

In all previous studies, authors have relied on large-$\mathcal{N}$ approximations in order
to calculate the gravitational wave signal.  Here we make no such approximation
and approach the problem numerically.  We place $\mathcal{N}$ scalar fields on a lattice evolving in a
Friedmann-Lema\^itre-Robertson-Walker background.  Each lattice point is
initialized with a field value derived from thermal initial conditions.   If we
allow the lattice spacing to be $H_0^{-1}$ at the beginning of the simulation
each lattice point will settle into an independent position in the vacuum
manifold.  As the simulation progresses, the fields will evolve and align
themselves.  This self-ordering produces anisotropic stresses that source 
gravitational waves.

This paper is organized as follows.  In Section~\ref{GSB} we will introduce a toy
model of spontaneous symmetry breaking.  We outline our computational
methods in Section~\ref{GWP} and present the spectra we calculate in our
simulations.  In Section~\ref{discussion} we will comment on the differences 
between the spectra we predict from self-ordering and the spectrum 
produced during inflation.

\section{Global Symmetry Breaking}
\label{GSB}

The GUT scale naturally arises from particle physics as 
the scale at which the electroweak and strong couplings are of the same order 
of magnitude.  It is likely that these forces are unified under some larger symmetry 
above the GUT scale and that this symmetry is broken as the Universe cools.
Nevertheless, the nature of the GUT symmetry and how it is broken remains unknown.

The transition could be first-order, in which case bubbles of the broken phase 
nucleate and coalesce.  In this model the phase transition happens very 
rapidly---the entirety of the universe can end up in a unique state in less than
a Hubble time.  This process is likely to produce gravitational radiation 
\cite{Witten:1984rs,Kosowsky:1991ua,Kamionkowski:1993fg,Child:2010} as bubbles collide 
and coalesce.

Conversely, the phase transition could be second-order.  In this case the field
smoothly transitions to the broken phase as the temperature of the universe drops. 
If the broken phase is not unique, that is
if the vacuum state has some symmetry with respect to the field configuration, 
the effects of the {\sl existence} of this phase transition can lead to observational 
effects for many Hubble times.

We begin with two assumptions: (1) that the Universe is radiation dominated
at the time when the phase transition occurs, and (2) that the energy associated with the
fields undergoing the phase transition is some small fraction, $\alpha$, of the 
total energy density at the time of the transition, $\rho_c$.  The total 
density, at any time, is
\begin{equation}
\rho = \rho_{\rm rad} + \rho_{\phi}
\end{equation}
where
\begin{equation}
\label{tdeppot}
\rho_\phi = \sum_i \frac{1}{2}\left(\dot{\phi}^2 + \frac{\left(\nabla \phi_i\right)^2}{a^2}\right) + V(\phi_i,T)
\end{equation}
and 
\begin{equation}
\rho_{\rm rad} = (1-\alpha)a^{-4}\rho.
\end{equation}
Since the Universe is necessarily dominated 
by the radiation energy-density, we will only consider cases where $\alpha \ll 1$ so
that the Universe is well described by assuming $H\propto a^{-4}$.   

The potential in \Deqn{tdeppot} is temperature dependent.  To leading order in
temperature, 
\begin{equation}
\label{tpot}
V(\phi_i,T) = m^2_{\rm eff}(T)\phi^2 +
\frac{\lambda}{8}\left(\phi^4+\frac{v^4}{4}\right),
\end{equation}
where $\phi^2 = \sum_i \phi_i^2$.  The temperature dependent effective mass can 
be parameterized by
\begin{equation} 
m_{\rm eff}^2 = \frac{\lambda v^2}{8} \left(\frac{T}{T_c}-1\right).
\end{equation} 
At temperatures higher than the
critical temperature, $T_c$, the effective mass is positive, the potential has 
a unique minimum at the origin, and this minimum has full $O(\mathcal{N})$ 
symmetry, and at the origin 
\begin{equation} 
\left.m_{\rm eff}^2\right|_{\phi=0}= - \frac{\lambda v^2}{8}.
\end{equation} 
After the phase transition the potential has
an $O(\mathcal{N}-1)$ symmetric VEV
\begin{equation} 
\label{potential}
V(\phi_i) = V(\phi_i,0) = \frac{\lambda}{8} \left(\phi^2-\frac{v^2}{2}\right)^2,
\end{equation}

The phase transition occurs at the critical temperature, $T_c$, when
the effective mass of the field vanishes.
Although the field has a mean value, $\phi=0$, there is a variance associated
with this value, 
\begin{equation}
\sigma^2 = \left\langle\phi^2\right\rangle -\left\langle\phi \right\rangle^2 =  \left\langle\phi^2\right\rangle,
\end{equation}
that sets the distribution of field values at the time of the transition.  We can assume 
that at this time, each Hubble volume
(sphere of radius
$H_c^{-1}$) has a homogeneous field value, drawn from a Gaussian distribution (see
Appendix~\ref{thermal}) 
\begin{equation} 
P(\phi) = \sqrt{\frac{1}{2\pi \sigma^2}}e^{-\frac{\phi^2}{2\sigma^2}}, 
\end{equation}
and
\begin{eqnarray} 
\nonumber
\sigma^2 &=& \frac{T_c}{2\pi^2 }\int_0^\infty \frac{k^2e^{- k^2/H_c^2}}{k^2+m_{\rm eff}^2}
dk \\ &=& \frac{H_cT_c}{4\pi^{3/2}}.\\ \nonumber 
\end{eqnarray} 
where the second
equality comes from setting $m_{\rm eff}=0$.

The temperature of the Universe at the beginning of the simulation, $T_c$, is related to 
the energy density of the Universe at that time,
\begin{equation}
\rho_c = \frac{\pi^2}{30}g_c T_c^4,
\end{equation}
where $g_c$ is the number of ultra-relativistic degrees of freedom at the time of the 
phase transition.  We take $g_c = 1000$.

The average energy density in the field is well approximated by setting $\phi=0$ in \Deqn{potential},
\begin{equation} 
\left<\rho_\phi \right> \approx \frac{\lambda v^4}{32},
\end{equation}
which is some fraction, $\alpha$, of the total energy density 
\begin{equation}
\frac{\lambda v^4}{32}  = \alpha \rho_c = \alpha \frac{3m_{\rm pl}^2}{8\pi} H_c^2.
\end{equation} 
This constrains the value of the VEV, 
\begin{equation}
\label{defofvev}
\frac{v^2}{2} =  \sqrt{\frac{3\alpha}{\lambda \pi}}H_cm_{\rm pl}.
\end{equation}
This provides us with a good self-consistency check, namely, that the variance
of the fluctuations of the filed are small compared to the VEV, $v/\sqrt{2}$.
 This ratio is given by 
\begin{equation} 
\frac{\sigma^2}{v^2/2} =
\frac{T_0}{m_{\rm pl}}\frac{1}{4\pi}\sqrt{\frac{\lambda}{3\alpha}},
\end{equation}
which is, in general, less that one if $T_0$ is significantly below $10^{19}\,{\rm GeV}$ 
and $\alpha$ and $\lambda$ are of similar order.
\section{Gravitational Waves}
\label{GWP}

Evolving scalar fields on a discrete lattice is now a mature field of study.  Scalar fields were first introduced to the lattice, in a Cosmological context, by {\sc LatticeEasy}~\cite{Felder:2000hq} and later  by {\sc DeFROST} \cite{Frolov:2008hy} to study the non-linear dynamics of preheating after inflation.  More recently the authors of \cite{Easther:2010qz} re-framed the question by moving the fundamental description of the fields from configuration space to momentum space.   Even more recently the author of \cite{Huang:2011gf} introduced a versatile code that allows the user more control over the integrating scheme.  Here we chose to use {\sc LatticeEasy} since we are interested in sub- and super-horizon scales, large lattices with efficient storage and a specific associated potential, \Deqn{potential}. This software natively evolves scalar fields  according to the Klein-Gordon equation in an expanding background, 
\begin{equation} 
\ddot{\phi}_i + 3H \dot{\phi}_i -
\frac{\nabla^2\phi_i}{a^2} + \frac{\partial V(\phi_i)}{\partial \phi} = 0,
\end{equation} 
where we  work in units where $c=\hbar=1$.  The homogeneous background evolution is determined by 
\begin{equation}
\label{firstfried}
H^2 = \frac{8\pi}{3m_{pl}^2}\rho,
\end{equation}
where $\rho=\rho(t)$ is the homogeneous, average, energy density at time $t$.  We couple {\sc LatticeEasy} to a code that evolves the metric perturbation using the methods
of~\cite{Easther:2006vd,Easther:2007vj}.

Since the lattice realizes the fields at discrete values of time, it is most convenient to perturb the metric, $h_{ij}$, in a synchronous gauge 
\begin{equation} 
\label{sync} ds^2 = dt^2 - a^2(t)\left[\delta_{ij} +
h_{ij}\right]dx^idx^j.  
\end{equation} 
Additionally, the radiative part of $h_{ij}$ obeys the transverse-traceless conditions 
\begin{equation}
h_i^i = 0\,\,\,{\rm and} \,\,\,\,h_{ij,j} =0.  
\end{equation} 
The radiative perturbations obey sourced Klein-Gordon equations 
\begin{equation} 
\ddot{h}_{ij} + 3H \dot{h}_{ij} - \frac{1}{a^2} \nabla^2 h_{ij} = \frac{16 \pi}{m^2_{\rm pl}} S^{TT}_{ij},
\label{equationofmotion} 
\end{equation} 
where the source term, $S^{TT}_{ij}$ is
the transverse-traceless projection of the anisotropic stress tensor, 
\begin{equation}
\label{aniso} S_{ij} = T_{ij} - \frac{\eta_{ij}}{3}T.  
\end{equation} 

We specify our model \Deqn{potential} and allow {\sc LatticeEasy} to evolve the
fields and the scale factor.  We can then calculate the source term of
\Deqn{equationofmotion}, and evolve the six metric perturbations, $h_{ij}^{TT}$.
We can always check our numerical stability by checking to see if the metric
perturbations are still transverse-traceless;  transverse-traceless metric perturbations require {\sl both} a transverse-traceless source and accurate evolution.

At any point during the simulation, we can calculate the power spectrum of gravitational radiation.  The stress-energy associated with metric perturbations is \cite{Misner:1974qy}, 
\begin{equation} 
T^{\rm
gw}_{\mu\nu} = \frac{1}{32\pi} \left\langle h_{ij,\mu}h^{ij}_{\,\,\,,\nu}\right\rangle,
\end{equation} 
where the brackets denote a spatial average over at least a few wavelengths.  The $00$ component is the energy density,
\begin{equation} 
\rho_{\rm{gw}} =
\frac{t^{\mu}t^{\nu}}{32\pi} \left\langle h_{ij,\mu}h^{ij}_{,\nu}\right\rangle = \frac{1}{32
\pi} \sum_{i,j} \left\langle\dot{h}^2_{ij}\right\rangle, \label{gwdensity} 
\end{equation}
where $t^{\mu} = (1,0,0,0)$. Finally, we can invoke Parseval's theorem (see
\cite{Easther:2007vj}) to rewrite \Deqn{gwdensity} as
\begin{equation} 
\rho_{\rm gw} =\frac{1}{32\pi}
\frac{1}{V}\sum_{i,j}\int d^3\mathbf{k}\,\,
\Bigl|\dot{h}_{ij}(t,\mathbf{k})\Bigr|^2, \label{omega0} 
\end{equation} 
where $V$ is the comoving volume over which the spatial average is being performed. We can
then write 
\begin{equation} 
\frac{d\rho_{\rm gw}}{d\ln k} = \frac{k^3}{32\pi}
\frac{1}{V} \sum_{i,j} \int d\Omega\, \Bigl| \dot{h}_{ij}^{\rm
TT}(\eta,\mathbf{k}) \Bigr|^2, \label{omega} 
\end{equation} 
which can be transferred to present-day amplitude and frequency by ~\cite{Easther:2007vj,Price:2008hq},
\begin{equation} 
\Omega_{\rm gw,0}h^2 = \Omega_{\rm rad,0}h^2
\Biggl(\frac{g_0}{g_e}\Biggr)^{1/3} \frac{1}{\rho_{\rm tot, e}}\frac{d\rho_{\rm
gw,e}}{d\ln k}, \label{omega1} 
\end{equation} 
where the $0$ and $e$ subscripts denote
quantities defined today and the end of our simulations, respectively.  We also keep the convention that $h$ absorbs the uncertainty in the present value of the
Hubble parameter, $\Omega_{\rm rad,0}$ is the current fraction of the energy
density in the form of radiation, and $\rho_{\rm tot, e}$ is the total energy
density at the end of our simulations. The ratio, $g_0/g_{\rm e}$, is the number
of degrees of freedom today to the number of degrees of freedom at
matter/radiation equality.  We approximate $g_0/g_{\rm e}=1/100$.

\section{Results}
\label{discussion}

The first major difference between the {\sl structure} of the gravitational-wave
spectrum from self-ordering and that predicted by inflation is the lack of power
at high-frequencies.  This cut-off feature exists because we only considering larger then Hubble length fluctuations as sources of gravitational waves as there will be no/short lived gradient terms to source the gravitational waves inside the Hubble volumes by second order phase transitions.
\begin{figure}[htbp]
   \centering
   \includegraphics[width=3in]{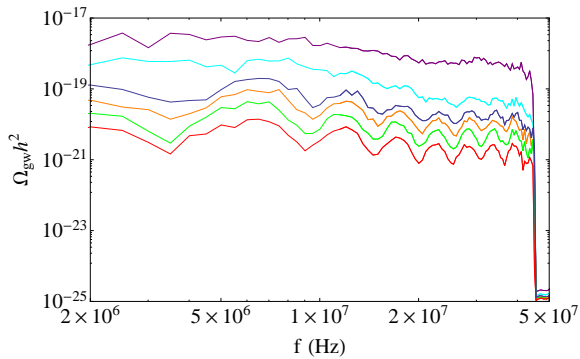} 
   \caption{The present-day gravitational wave spectrum from self-ordering.  From top to bottom, $\mathcal{N}=2,3,4,5,8,16$.}
   \label{fig:ncomparison}
\end{figure}

\begin{figure}[htbp]
   \centering
   \includegraphics[width=3in]{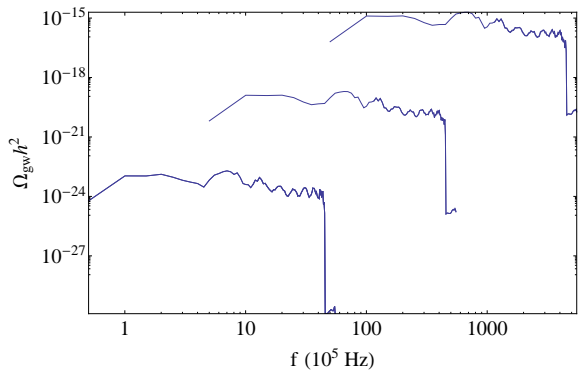} 
   \caption{The present-day gravitational wave spectrum from self-ordering.  From top (rightmost) to bottom (leftmost), $\rho_c^{1/4}=10^{-3}\,m_{\rm pl}$, $\rho_c^{1/4}=10^{-4}\,m_{\rm pl}$, $\rho_c^{1/4}=10^{-5}\,m_{\rm pl}$.}
   \label{fig:soucomparison}
\end{figure}

We see in \cite{Easther:2006gt} that the cut-off frequency is related to the Hubble length at the time when the gravitational wave is generated
\begin{equation}
f_{\rm peak} = 6\times 10^{-10} \frac{k}{\sqrt{m_{\rm pl} H}}\,\rm{Hz}.
\end{equation}
This, along with and the first Friedmann equation,
\begin{equation}
H_c = \sqrt{\frac{8\pi}{3}}\frac{\sqrt{\rho_c}}{m_{\rm pl}},
\end{equation}
allows us to determine where the cut-off should appear
\begin{equation}
\label{peakeq}
f_{\rm peak} \sim 10^{11} \frac{\rho_c^{1/4}}{m_{\rm pl}}\,{\rm Hz}.
\end{equation}
For example, with $\rho_c^{1/4} = 10^{-4} m_{\rm pl}$ we expect the cutoff to be at $f = 10^{7}\, {\rm Hz}$ which agrees with the results of our simulation, shown in \Dfig{fig:ncomparison}.  The characteristic amplitude and scale invariant nature of the spectrum was expected from the analytical approach of \cite{JonesSmith:2007ne,Krauss:2010df} and \cite{Fenu:2009qf}. It is interesting to note that the spectrum is scale invariant for all the cases in \Dfig{fig:ncomparison}, including those for $\mathcal{N}=2,3$; the analytical  methods employed by \cite{Fenu:2009qf} assumed that $\mathcal{N}$ is large.

In Fig.~\ref{fig:soucomparison} we observe two important scaling effects.  First, we recover the fact that the high-frequency cutoff predicted by Eq.~\ref{peakeq} scales with $\rho^{1/4}$.  We also see that the amplitude of the signal is proportional to the energy density of the Universe at the time of the transition,
\[\Omega_{\rm gw}h^2 \propto \rho_c.\]
This scaling is also suggested predicted in \cite{JonesSmith:2007ne,Krauss:2010df} and \cite{Fenu:2009qf}.

A final step is to compare the numerical results here with the analytic arguments of \cite{JonesSmith:2007ne,Krauss:2010df} and \cite{Fenu:2009qf}.  In both cases the authors use the model presented here to estimate the gravitational wave signal from reordering.  The two sets of authors use slightly different parameterizations of the model; however, all authors arrive at the conclusion that there should be a scale-invariant gravitational wave spectrum from this transition

In \cite{JonesSmith:2007ne,Krauss:2010df}, the authors estimate the power in
gravitational waves from field reordering to be (Eq.~(10) in \cite{Krauss:2010df})
\begin{equation}
\label{krasseq}
\Omega^{\rm JKM}_{\rm gw}= \frac{1}{12\pi^3H_0^2}k^2 P(k,\tau_0),
\end{equation}
where $P(k,\tau)$ is calculated numerically,  the Hubble constant today is $H$, $\tau_0$ is conformal time today and we've modified the normalization constant $1/12\pi^3$ to reflect a convention choice in \cite{JonesSmith:2007ne} and a typographical error in \cite{Krauss:2010df}.  The strain power, $P$, ends up being a function only of $k\tau$.  It peaks around $k\tau \approx 3.7$, although this corresponds to modes that entered the horizon during matter domination.  To ensure that we're identifying modes that entered during radiation domination, we choose a mode, $k\approx 10/\tau_{\rm eq}$, 
\begin{equation}
\label{maxpowerkrauss}
P\left(\frac{k}{\tau_{\rm eq}},\tau_0\right) = \Omega_{\rm rad} P\left(\frac{k}{\tau_{\rm eq}},\tau_{\rm eq}\right),
\end{equation}  
where we note that gravitational waves scale as a constant fraction of $\Omega_{\rm rad}$ after matter-radiation equality, the subscript ${\rm eq}$ indicates evaluating quantities at the time of radiation-matter equality.  We can read off $P(k/\tau_{\rm eq},\tau_{\rm eq})$ from  Fig.~1 in \cite{JonesSmith:2007ne},
\begin{equation}
P\left(\frac{k}{\tau_{\rm eq}},\tau_0\right) \approx 1000 \Omega_{\rm rad}.
\end{equation}
So we can estimate the total gravitational wave energy per octave, 
\begin{equation}
\label{estimate2}
\Omega^{\rm JKM}_{\rm gw}=  \frac{1}{12\pi^3H_0^2}\frac{10^2}{\tau_{\rm eq}^2}a_{\rm eq}^2 P\left(\frac{3.7}{\tau_{\rm eq}},\tau_{\rm eq}\right),
\end{equation}
where the physical wavevector now is $1/a_{eq}$ larger than the physical wavevector at the time of radiation-matter equality.  We can make a crude estimate of the value of conformal time at radiation-matter equality,
\begin{equation}
\label{conftimenow}\tau_{\rm eq} = \frac{1}{H_{\rm eq}}\int_0^{a_{\rm eq}} da^\prime \left(\Omega_R+\Omega_Ma+\Omega_\Lambda a^4\right)^{-1/2}\sim\frac{1}{50H_0},
\end{equation}
where the fractional energy densities come from \cite{Komatsu:2010fb}. Putting \Deqn{maxpowerkrauss} together with Eqs.~(\ref{estimate2},\ref{conftimenow}), we get an estimate,
\begin{equation}
\label{Jestimate}
\Omega^{\rm JKM}_{\rm gw}h^2= \frac{99}{\mathcal{N}}\Omega_{\rm rad}h^2
\left(\frac{v^4}{4\mathcal{N}m_{\rm pl}^4}\right), \\
\end{equation}
or, imposing our parameterization of the current-day Hubble constant, setting $\Omega_{\rm rad}h^2\approx 2\times10^{-5}$ and using Eqs.~(\ref{defofvev},\ref{firstfried})
\begin{equation}
\Omega^{\rm JKM}_{\rm gw}h^2 = \frac{0.016}{\mathcal{N}}
\frac{\alpha}{\lambda}\frac{\rho_c}{m_{\rm pl}^4}.
\end{equation}

In \cite{Fenu:2009qf}, the authors predict a scale-invariant power spectrum (Eq.~(5.2) of \cite{Fenu:2009qf}) 
\begin{equation}
\label{fenuresult}
\Omega^{\rm FFDG}_{\rm gw}h^2 \simeq \frac{511}{\mathcal{N}}\Omega_{\rm rad}h^2 \left(\frac{v}{\sqrt{2}m_{\rm pl}}\right)^4,
\end{equation}
where we use our definition of $v$ and our parameterization $\Omega_{gw}h^2$.  Using Eqs.~(\ref{defofvev},\ref{firstfried}) along with $\Omega_{\rm rad}h^2 \approx 2\times10^{-5}$, the expression in \Deqn{fenuresult} reduces to
\begin{equation}
\label{Festimate}
\Omega^{\rm FFDG}_{\rm gw}h^2 \simeq \frac{0.082}{\mathcal{N}}\frac{\alpha}{\lambda} \frac{\rho_c}{m_{\rm pl}^4}.
\end{equation}

These two estimates should vary from our simulations by one important factor.  In both cases, the Universe is comprised only of the scalar fields.  To preserve a radiation-dominated phase during and after the phase transition, we have, inherently, diluted the source by a factor of $\alpha$ which dilutes the analytic estimates Eqs.~(\ref{Jestimate},\ref{Festimate}) by a factor of $\alpha^2$.

It is worth pointing out that some of the phase transitions we have simulated 
result in the production of global
topological defects. Specifically, global strings for $\mathcal{N}=2$, 
global monopoles $\mathcal{N}=3$, and global textures for $\mathcal{N}>3$. 
Surprisingly, we find that the gravitational radiation produced 
is consistent with the large $\mathcal{N}$ approximation even for low values 
of $\mathcal{N}$, where the approximation is not valid (see our 
analytic estimates above). The gravitational wave backgrounds produced 
by global strings and monopoles are larger than those produced by textures.
We will investigate these cases in more detail in a future publication.

\section{Discussion}

Phase transitions at high energies are a generic consequence of (almost) all models of high energy physics.  Since there is no 
unique model of physics at this scale, we are forced to look for generic observational consequences at this scale.  Second order 
phase transitions will not produce gravitational radiation over one Hubble time, yet the reordering of the fields that mediate
 the transition could produce characteristic gravitational radiation over a wide range of scales.  Although this signature 
 could be misinterpreted as the gravitational radiation from primordial quantum fluctuations, it might be possible to distinguish the two at 
 very high frequencies.  Such a high-frequency detection would also carry information about the energy scale at which the phase transition
 occurred.

\begin{table}
  \caption{\label{tab:specN} Spectral amplitudes as a
function of number of fields for simulations with
$\left(\rho_c^{1/4}=10^{-4}m_{pl}, \alpha = \lambda = 0.1 \right)$.  The
numerical values, $\Omega_{\rm gw}h^2$, are an average value taken from the simulations, while the
 values in the second two columns are obtained from \Deqn{Jestimate} or \Deqn{Festimate}.} 
 \begin{tabular}{cccc}    
\hline
\hline
$\mathcal{N}$ & $\Omega_{\rm gw} h^2$ & $\alpha^2 \Omega^{\rm JKM}_{\rm gw} h^2$ & $\alpha^2 \Omega^{\rm FFDG}_{\rm gw}h^2$ \\
\hline
   2 & $1.0\times10^{-18}$ &$9.0\times10^{-21}$ & $4.1\times10^{-20}$ \\
   4 & $3.8\times10^{-20}$ & $4.0\times10^{-21}$& $2.1\times10^{-20}$ \\
   8 & $8.3\times10^{-21}$ & $2.0\times10^{-21}$& $1.0\times10^{-20}$ \\
  16 & $3.1\times10^{-21}$ & $1.0\times10^{-21}$& $5.1\times10^{-21}$ \\
 \end{tabular} 
\end{table}

\begin{table}
  \caption{\label{tab:specr} Spectral amplitudes as a
function of $\rho_c$ simulations with
$\left(\mathcal{N}=4, \alpha = \lambda = 0.1 \right)$.  The
numerical values, $\Omega_{\rm gw}h^2$, are an average value taken from the simulations, while the
 values in the second two columns are obtained from \Deqn{Jestimate} or \Deqn{Festimate}.} 
 \begin{tabular}{cccc}    
\hline
\hline
$\rho_c^{1/4}(m_{pl})$ & $\Omega_{\rm gw} h^2$& $\alpha^2 \Omega^{\rm JKM}_{\rm gw} h^2$ & $\alpha^2\Omega^{\rm FFDG}_{\rm gw} h^2$    \\
\hline
   $10^{-3}$ & $4.7\times10^{-16}$ & $4.0\times10^{-17}$ & $2.1\times10^{-16}$ \\
   $10^{-4}$ & $3.8\times10^{-20}$ & $4.0\times10^{-21}$& $2.1\times10^{-20}$ \\
   $10^{-5}$ & $4.0\times10^{-24}$ & $4.0\times10^{-25}$& $2.1\times10^{-24}$ \\
 \end{tabular} 
\end{table}

We considered the phenomenological model of
\cite{JonesSmith:2007ne,Krauss:2010df} and \cite{Fenu:2009qf} in which 
an $\mathcal{O(N)}$ symmetric false vacuum is dynamically broken into a
$\mathcal{O}\left(\mathcal{N} - 1 \right)$ true vacuum, we find this produces a scale
invariant gravitational wave spectrum whose amplitude depends inversely on
number of fields.  These results are summarized in Table~\ref{tab:specN}.   Our
numerical results suggest that the large $\mathcal{N}$ is not needed to
make a scale invariant spectrum.  Since the results of \cite{JonesSmith:2007ne,Krauss:2010df,Fenu:2009qf}
are derived using a large $\mathcal{N}$ approximation for the amplitude, one does not expect
these estimates to be a perfect estimator of the amplitude of the gravitational waves in the low-$\mathcal{N}$ limit.
Additionally, the results for varying values of $\rho_c$ are given in
Table~\ref{tab:specr}.  We find that our simulations differ from analytic estimates by only a small amount
and are more consistent at large $\mathcal{N}$, consistent with the fact that analytic methods are derived
from a large $ \mathcal{N}$ expansion.

\section{Acknowledgments}

We thank Latham Boyle, Andrew Tolley, Harsh Mathur and Katherine Jones-Smith  for useful 
discussions.  JTG is
supported by the National Science Foundation, PHY-1068080, and a Cottrell College Science 
Award from the Research Corporation. The work of XS and BV is supported by National Science 
Foundation grants PHY-0970074, PHY-0955929, 
and PHY-0758155, and the University of Wisconsin--Milwaukee Research Growth Initiative.
Research at the Perimeter Institute for Theoretical Physics is supported by the
Government of Canada through Industry Canada and by the Province of Ontario
through the Ministry of Research \& Innovation.  JTG would also like to thank
the University of Wisconsin-Milwaukee for its generous hospitality while some of
this research was completed.  LRP would like to thank the Perimeter Institute
for Theoretical Physics for their hospitality while some of this work was being
completed there.

\appendix

\section{Thermal Initial Conditions} \label{thermal}

The probability of the scalar field taking some average value, $\bar\phi$, is
given by 
\begin{equation} 
\label{fullprob}
P(\phi_v = \bar{\phi}) =  \int D\phi {\mathcal P}[\phi] \delta(\phi_v - \bar{\phi}) =
\langle \delta(\phi_v-\bar{\phi})\rangle,
\end{equation}
where ${\mathcal P}[\phi]$ is the probability functional and we've introduced
the volume-averaged field
\begin{eqnarray*}
\phi_v &=& \frac{1}{V}\int_V d^3x\,\phi(x) \\
&=& \frac{1}{V}\int_{-\infty}^{\infty} d^3x\, I(x)\phi(x),
\end{eqnarray*}
where $I(x)$ is a window function introduced for later computational
convenience.  We compute \Deqn{fullprob} using a Gaussian
approximation
\begin{equation*}
P(\phi_v = \bar{\phi}) \approx \sqrt{\frac{1}{2\pi \sigma^2}}
\exp{\left({-\frac{(\bar{\phi}-\mu)^2}{2 \sigma^2}}\right)},
\end{equation*}
with $\mu = \langle\phi_v\rangle$ and $\sigma^2 =
\langle\phi_v^2\rangle-\langle\phi_v\rangle^2$.  The problem is now to compute
the moments of the field
\begin{eqnarray}
\langle\phi_v\rangle &=& \int D\phi \mathcal{P}[\phi]\left(\frac{1}{V}\int_V
d^3x\, \phi(x) \right) \label{firstmoment}\\
\langle\phi_v^2\rangle &=& \int
D\phi \mathcal{P}[\phi] \left(\frac{1}{V^2}\int_V
d^3xd^3y\, \phi(x)\phi(y) \right)\label{secondmoment} 
\end{eqnarray}
The full probability functional can be found in, e.g.~
\cite{Hindmarsh:1993mb}.  Here we consider only the leading order temperature
dependence, which modifies the potential by replacing $m$ with a temperature
dependent term, $m_{\rm eff}(\beta)$, where $\beta\propto T^{-1}$.  Choosing a
temperature above $m_{\rm eff} = 0$ puts us in a symmetric phase of the
effective potential.  We then have 
\begin{eqnarray*} 
\mathcal{P}[\phi] &\approx& \frac{1}{Z} \exp{\left(-\beta H_{\mathrm{eff}}[\phi]
\right)}\\
&=&\frac{1}{Z} \exp{\left(\frac{-\beta}{2}\int d^3x\,  \phi(x)(\nabla +
m_{\mathrm{eff}}(\beta)^2)\phi(x)\right)}.
\end{eqnarray*}
Henceforth we write $m_{\rm eff}=m_{\rm eff}(\beta)$.  Defining
\begin{eqnarray*}
W[J] &=& \int D\phi \exp\Bigl(\frac{-\beta}{2}\int d^3x\, \phi(x)(\nabla +
m_{\rm eff}^2)\phi(x)\\
&\phantom{=}&\phantom{\int D\phi\exp -\beta\int d^3x} + J(x)\phi(x)\Bigr) \\
&=& Z \exp{\left(\frac{1}{2 \beta} \int \int d^3x d^3y J(x)
K^{-1}(x-y) J(y)\right)}, 
\end{eqnarray*}
with $ (-\nabla_x^2 + m_{\rm eff}^2)K^{-1}(x-y) = \delta^3(x-y)$ allows us to write the
moments of the field as
\begin{eqnarray*}
\langle\phi_v\rangle &=&  \frac{1}{V}\int_{-\infty}^{\infty} d^3x\,
I(x)\frac{\delta}{\delta J(x)}W[J]\Bigr|_{J=0} \\
\langle\phi_v^2\rangle &=&  \frac{1}{V}\int_{-\infty}^{\infty} d^3xd^3y\,
I(x)I(y)\nonumber\\
&\phantom{=}&
\phantom{\frac{1}{V}\int_{-\infty}^{\infty}}\times\frac{\delta}{\delta
J(x)}\frac{\delta}{\delta J(y)}W[J]\Bigr|_{J=0}.
\end{eqnarray*}
Choosing a Gaussian window such that $\int d^3x\, I(x)=4\pi R^3/3$ and evaluating the integrals leads to
\begin{eqnarray*}
\langle\phi_v\rangle &=& 0 \\
\langle\phi_v^2\rangle &=& \frac{1}{4 \pi^{3/2} \beta R} - \frac{m_{\rm eff} \exp\left(m_{\rm eff}^2
R^2\right)}{4 \pi \beta R} {\rm erfc}(m_{\rm eff} R).
\end{eqnarray*}
Putting this all together we have
\begin{equation} 
P(\phi_v = \bar{\phi}) = \sqrt{\frac{H_0 T_0}{ 8 \pi^{5/2}}} \exp{\left(-\frac{\bar{\phi}^2}{4 \pi^{3/2}} H_0 T_0\right)}.
\end{equation}

\end{document}